\documentclass[twocolumn,showpacs,preprintnumbers,amsmath,amssymb,nofootinbib]{revtex4-1}
\usepackage{graphicx}

\usepackage{color}
\usepackage[normalem]{ulem}  % \sout{old text} for strikeout

\renewcommand\sout{\bgroup \color{red}\ULdepth=-.5ex \ULset}
\newcommand\soutb{\bgroup \color{blue} \ULdepth=-.5ex \ULset}

\usepackage{graphicx}% Include figure files
\usepackage{dcolumn}% Align table columns on decimal point
\usepackage{bm}% bold math

\begin{document}

\preprint{}
\def\vc#1{\mbox{\boldmath $#1$}}

\title{Electric dipole moment of $^{13}$C
}% Force line breaks with \\

\author{Nodoka~Yamanaka$^{1,2}$}
  \email{nodoka.yamanaka@riken.jp}
\author{Taiichi~Yamada$^3$}
\author{Emiko~Hiyama$^4$}
\author{Yasuro~Funaki$^{5}$}
\affiliation{$^1$IPNO, Universit\'{e} Paris-Sud, CNRS/IN2P3, F-91406, Orsay, France}
\affiliation{$^2$iTHES Research Group, RIKEN, %\\
Wako, Saitama 351-0198, Japan}
\affiliation{$^3$Laboratory of Physics, Kanto Gakuin University,%\\
Yokohama 236-8501, Japan}
\affiliation{$^4$RIKEN Nishina Center, RIKEN, %\\
Wako, Saitama 351-0198, Japan}
\affiliation{$^5$School of Physics and Nuclear Energy Engineering and IRCNPC,
Beihang University, Beijing 100191, China}

\date{\today}% It is always \today, today,
             %  but any date may be explicitly specified

\begin{abstract}

We calculate for the first time the electric dipole moment (EDM) of $^{13}$C generated by the isovector CP-odd pion exchange nuclear force in the $\alpha$-cluster model, which describes well the structures of low lying states of the $^{13}$C nucleus.
The linear dependence of the EDM of $^{13}$C on the neutron EDM and the isovector CP-odd nuclear coupling is found to be $d_{^{13}{\rm C}} = -0.33 d_n - 0.0020 \bar G_\pi^{(1)}$.
The linear enhancement factor of the CP-odd nuclear coupling is smaller than that of the deuteron, due to the difference of the structure between the $1/2^-_1$ state and the opposite parity ($1/2^+$) states.
We clarify the role of the structure played in the enhancement of the EDM.
This result provides good guiding principles to search for other nuclei with large enhancement factor.
We also mention the role of the EDM of $^{13}$C in determining the new physics beyond the standard model.

\end{abstract}

% PACS, the Physics and Astronomy
                             % Classification Scheme.
%CP invariance, 11.30.Er
%Electric moments, nuclear, 21.10.Ky
%Nuclear tests of fundamental interactions and symmetries, 24.80.+y
%Cluster model, nuclear structure, 21.60.Gx

%\keywords{Suggested keywords}%Use showkeys class option if keyword
                              %display desired
\maketitle

%%%%%%%%%% Introduction %%%%%%%%%%%%%%%

\section{Introduction\label{sec:intro}}

One of the most important challenge in particle physics and cosmology is to explain the baryon number asymmetry of the Universe.
In the standard model, it is however known that the CP violation is not sufficient to realize the currently observed matter abundance \cite{sakharov,shaposhnikov1,shaposhnikov2,farrar,huet}.
This actually gives a strong motivation to search for new physics beyond the standard model with large CP violation.

The most promising experimental observable to probe the CP violation of new physics is the {\it electric dipole moment} (EDM) \cite{hereview,Bernreuther,khriplovichbook,ginges,pospelovreview,fukuyama,naviliat,engel,yamanakabook,hewett,roberts}.
The EDM is measurable in many systems, and many experimental measurements were done so far, including the neutron \cite{baker}, atoms \cite{rosenberry,regan,griffith,graner,parker}, molecules \cite{hudson,acme}, etc.
Recently, new technology using storage rings is making it possible to measure the EDM of charged particles \cite{muong-2,storage1,storage2,storage3,storage4,storage5,storage6,storage7,bnl}.
The experimental measurement of the nuclear EDM is also developed, and is expected to be realized in the near future.

The study of the nuclear EDM has several notable advantages.
The first advantage is the absence of electrons which screen the EDM of the nucleus, as dictated by the theorem of Schiff \cite{schiff}.
The second one is the small contribution from the standard model CP violation \cite{ckm}, which makes it an experimentally clean observable \cite{smtheta,czarnecki,smneutronedmmckellar,mannel,seng,yamanakasmedm}.
The final reason is the potential enhancement of the EDM by the nuclear many-body effect \cite{sushkov,sushkov2,sushkov3}.

Due to those advantages, theoretical evaluations of the nuclear EDM have been extensively done so far \cite{avishai1,avishai2,avishai3,korkin,pospelovdeuteron,liu,afnan,devries,dedmtheta,stetcu,chiral3nucleon,song,dekens,bsaisou,bsaisou2,mereghetti,devries2,devries3,yoshinaganuclearedm,yamanakanuclearedm}.
The EDM of nuclei was first estimated in the simple core+valence model, and an enhancement of the CP violation was suggested \cite{sushkov,sushkov2,sushkov3}.
For light nuclei, the applicability of this model is however questionable, since other correlations such as the cluster structure \cite{clusterreview1,clusterreview2,clusterreview3}, which are thought to be important in light nuclear systems, were not taken into account.
In a recent work, the EDM of light nuclei was evaluated in the cluster model, and it was found that the cluster structure may enhance the nuclear EDM \cite{yamanakanuclearedm}.
There the EDMs of $^6$Li and $^9$Be were calculated within the framework of $\alpha +p + n$ and $2\alpha +n$ three-body systems, respectively.
The cluster model was successfully applied in the calculation of the $^6$Li and $^9$Be nuclei, and the result suggested an enhancement compared to the deuteron EDM, thanks to the cluster structure.
This fact motivates us to theoretically evaluate the EDM of other unknown light nuclei with the expectation to find sensitive and experimentally advantageous nuclear systems.

The study of the EDM of multiple hadronic systems is not only essential in finding experimentally sensitive observables, but it is also absolutely necessary for constraining the hadron level CP violating interaction with many unknown couplings.
Recent investigations of hadronic CP violation in the chiral effective field theory is indicating that some CP violating interactions such as the three-pion interaction or the contact CP-odd nuclear force also receive contribution from the quark and gluon level CP violations in the leading order \cite{chiral3nucleon,devries3,eft6dim}.
As the determination of unknown couplings requires at least the same number of experimental values of different observables, further study of the EDM of light nuclei is mandatory to unveil the CP violation beyond the standard model.

As mentioned, the EDM of light nuclei has been studied up to the three-body system in the $\alpha$-cluster model.
The next step is then to study four-body systems. 
In this work, we therefore evaluate the EDM of $^{13}$C, which is the simplest one.
The $^{13}$C nucleus was recently studied within the framework of $3\alpha + n$ four-body cluster model \cite{yamada}.
The four-body problem can be solved by using the Gaussian expansion method \cite{hiyama,Hiyama2012ptep}, which was applied in many systems of a wide range of physical hierarchy \cite{kamimuramuonic,
3nucleon,
hiyamahypernuclei1,
hiyamahypernuclei2,
benchmark,
hiyamahyperon-nucleon,
hiyama1stexcitedstate,
yamadaschuck,
hiyamapentaquark,
hamaguchi,
funaki,
hiyamahypernuclei3,
hiyamaatom1,
hiyamaatom2,
ohtsubo,
yokota,
kusakabe,
yamada,maeda}.
In the study of $^{13}$C, its structure up to the excitation energy 15 MeV was successfully described, and
it was argued that this nucleus does not have a simple shell structure, so a significant change of the nuclear EDM from the simple core+valence model prediction is expected.
In this work, we will therefore calculate the nuclear EDM of $^{13}$C in the four-body cluster model using the Gaussian expansion method, and investigate the effect of the cluster correlations on the EDM.

The plan of this paper is as follows.
In the next section, we first give a brief overview of the calculation of the structure of $^{13}$C.
In Section \ref{sec:nuclearedm}, the formulation of the EDM and the CP-odd nuclear force is given.
We then show the result of the calculation of the EDM of $^{13}$C in the Gaussian expansion method, and analyze it.
In Section \ref{sec:roles}, we discuss the role of $^{13}$C EDM in the determination of hadron level CP violation.
In Section \ref{sec:newphys}, we show the prospects for the discovery of new physics beyond the standard model by measuring the EDM of $^{13}$C.
The final section is devoted to the summary.

\section{\label{sec:structure}Structure of $^{13}$C in the four-body cluster model}

Here we review the setup of the calculation of $^{13}$C and the CP-even nuclear force used within the framework of the $3\alpha$+$N$ four-body cluster model \cite{yamada}.
The total wave function of $^{13}$C $\tilde{\Psi}_J(A=13)$, with the total angular momentum $J$ and total isospin $T=\frac{1}{2}$ in the  $3\alpha+N$ cluster model is expressed by the product of the internal wave functions of $\alpha$ clusters $\phi(\alpha)$ and the relative wave function  $\Psi_J(A=13)$ among the $3\alpha$ clusters and the extra neutron,
\begin{eqnarray}
\tilde{\Phi}_J(A=13) = \Phi_J(A=13) \phi^{\rm}(\alpha_1) \phi(\alpha_2) \phi(\alpha_3).
\end{eqnarray}
The relative wave function $\Phi_J(A=13)$ is expanded in terms of the Gaussian basis as follows:
\begin{eqnarray}&&\Phi_J(A=13)= \sum_{p=1}^{4} \sum_{c^{(p)}} \sum_{\nu^{(p)}}f^{(p)}_{c^{(p)}}(\nu^{(p)}) \Phi^{(p)}_{c^{(p)}}(\nu^{(p)}), 
\label{eq:total_wf} \\
&&\Phi^{(p)}_{c^{(p)}}(\nu^{(p)}) = \mathcal{S}_{\alpha} \Bigl[ \Bigl[ \varphi_{\ell^{(p)}_1}(\vc{R}^{(p)}_1,\nu^{(p)}_1) \Bigl[ \varphi_{\ell^{(p)}_2}(\vc{R}^{(p)}_2,\nu^{(p)}_2)
\nonumber\\
&& \hspace{8em}
\varphi_{\ell^{(p)}_3}(\vc{R}^{(p)}_3,\nu^{(p)}_3) \Bigr]_{L^{(p)}_{23}} \Bigr]_{L^{(p)}} \xi_{\frac{1}{2}}(N) \Bigr]_{J},
\nonumber\\
\label{eq:total_wf_basis}\\
&& {\langle u_F | \Phi_J({^{13}{\rm C}}) \rangle } = 0, \label{eq:oc}
\end{eqnarray}
where we assign the cluster number, 1, 2, and 3, to the three $\alpha$ clusters (spin 0), and the number 4 to the extra nucleon (spin $\frac{1}{2}$). 
$\Phi^{(p)}_{c^{(p)}}(\nu^{(p)})$ denotes the relative wave function with respect to the $p$-th Jacobi-coordinate system of the four-body $3\alpha+N$ model (either of coordinate type of {\it K} or {\it H}) shown in Fig.~1 of Ref.~\cite{yamada}, where $c^{(p)}$ represents the angular momentum channel for the $p$-th Jacobi-coordinate system. 
$\mathcal{S}_{\alpha}$ stands for the symmetrization operator acting on all $\alpha$ particles obeying Bose statistic, and $\xi_{\frac{1}{2}}(N)$ is the spin function of the extra nucleon. 
$\phi^{\rm}(\alpha)$ denotes the intrinsic wave function of the $\alpha$ cluster with the $(0s)^{4}$ shell-model configuration. 
$\nu^{(p)}$ denotes the set of size parameters, $\nu^{(p)}_1$, $\nu^{(p)}_2$, and $\nu^{(p)}_3$, of the normalized Gaussian function, $\varphi_\ell(\vc{R},\nu)=N_\ell(\nu)R^\ell \exp(-\nu R^2) Y_\ell (\hat{\vc{R}})$, and $\nu$ is taken to be of geometrical progression,
\begin{equation}
\nu_n=1/b_n^2,\hspace{1em}b_n=b_{\rm min}a^{n-1},\hspace{1em}n=1\sim n_{\max}.
\label{eq:para_Gaussian} 
\end{equation}
It is noted that this prescription is found to be very useful in optimizing the ranges with a small number of free parameters ($b_{\rm min}$, $a$, $n_{\rm max}$) with high accuracy~\cite{kamimuramuonic,hiyama}.
In this work, the geometric series for each coordinate is made of seven terms, i.e. $7^3 = 343$ bases per angular momentum channel.
To converge the $1/2^-_1$ and $1/2^+_1$ states, we use 59  and 69 channels, respectively.
The calculation of the EDM requires parity mixing, so 128 angular momentum channels, i.e. 43904 bases are required.

The $3\alpha+N$ Hamiltonian for $\Phi_J(A=13)$ is presented as
\begin{eqnarray}
\mathcal{H}
&=&
\sum_{i=1}^{4} T_i-T_{\rm cm} + \sum_{i<j=1}^{3} V_{{2\alpha}}(i,j) + \sum_{i=1}^{3} V_{{\alpha}N}(i,4)
\nonumber \\
           && +V_{3\alpha}(1,2,3) + \sum_{i<j=1}^{3} V_{2\alpha N}(i,j,4) 
\nonumber\\
&& 
+ V_{3\alpha N}(1,2,3,4) + V_{\rm Pauli},
\label{eq:hamiltonian}
\end{eqnarray}
where $T_i$, $V_{2\alpha}$ ($V_{\alpha N}$), $V_{3\alpha}$ ($V_{2\alpha N}$), and $V_{3\alpha N}$ stand for the kinetic energy operator for the {\it i}-th cluster, $\alpha-\alpha$ ($\alpha-N$) potential, three-body potential among the three $\alpha$ particles (the two $\alpha$ particles and extra nucleon), and the four-body potential, respectively. The center-of-mass kinetic energy ($T_{\rm cm}$) is subtracted from the Hamiltonian. 

The effective $\alpha-\alpha$ potential $V_{2\alpha}$ is constructed by the folding procedure from an effective two-nucleon force including the proton-proton Coulomb force. 
Here we take the Schmid-Wildermuth (SW) force~\cite{schmid} as the effective $NN$ force. 
This folded $\alpha-\alpha$ potential reproduces nicely the $\alpha-\alpha$ scattering phase shifts and the energies of the $^8$Be ground-band state ($J^{\pi}=0^{+}-2^{+}-4^{+}$) within the framework of the $2\alpha$ cluster model. 
As for the $\alpha-N$ potential, we use the Kanada-Kaneko potential \cite{kanada}, which reproduces the low-energy $\alpha-N$ scattering phase shifts. 
We introduce the phenomenological effective three-body forces, $V_{3\alpha}$ and $V_{2\alpha N}$, the former depending on the total angular momentum of $^{12}$C.
The $3\alpha$ cluster model then describes well the structure of the low-lying states of $^{12}$C including the $2^{+}_2$, $0^{+}_3$, and $0^{+}_4$ states above the Hoyle state, which have quite recently been observed in experiments~\cite{freer,itoh}.
The structures of the low-lying states of $^9$Be and $^9$B are also well described by the $2\alpha+N$ cluster model. 
The four-body interaction, $V_{3\alpha N}$, is given to reproduce the $^{13}$C ground state energy. 

Equation (\ref{eq:oc}) represents the orthogonality condition that the total wave function (\ref{eq:total_wf}) should be orthogonal to the Pauli-forbidden states of the $3\alpha+N$ system, $u_F$'s, which are constructed from the Pauli forbidden states between two $\alpha$ particles and those between $\alpha$ particle and extra nucleon $N$~\cite{ocm1,ocm2,horiuchi1,ocm3,horiuchi2}.
The Pauli-forbidden states are removed by using the Pauli-blocking operator $V_{\rm Pauli}$~\cite{Kukulin} in Eq.~(\ref{eq:hamiltonian}),
\begin{eqnarray}
V_{\rm Pauli} = \lim_{\lambda \rightarrow  \infty}\ {\lambda}\ \sum_{f}\ {| u_f \rangle}{\langle u_f |},
\end{eqnarray} 
which rules out the Pauli-forbidden $\alpha$-$\alpha$ relative states ($f=0S,1S,0D$) and the Pauli-forbidden $\alpha-n$ relative state ($f=0S$) from the four-body $3\alpha-n$ wave function.
In the present study, we take $\lambda=10^{4}$~MeV.
The ground state of $^{13}$C with the dominant shell-model-like configuration $(0s)^{4}(0p)^{9}$, then, can be properly described in the present $3\alpha+N$ cluster model.

The equation of motion of $^{13}$C with the $3\alpha+N$ four-body cluster model is obtained by the variational principle,
\begin{eqnarray}
\delta\left[\langle \Phi_J(A=13)\mid \mathcal{H}-E \mid \Phi_J(A=13)\rangle\right]=0,
\label{eq:variational_principle}
\end{eqnarray}
where $E$ denotes the eigenenergy of $^{13}$C measured from the $3\alpha+n$ threshold.
The energy $E$ and the expansion coefficients $f^{(p)}_{c^{(p)}}$  in the total wave function shown in Eq.~(\ref{eq:total_wf}) are determined by solving a secular equation derived from Eq.~(\ref{eq:variational_principle}).

\section{\label{sec:nuclearedm}Calculation of the nuclear EDM}

To induce the nuclear EDM, the existence of the P, CP-odd nucleon level processes is required.
In this work, we assume the following effective CP-odd lagrangian \cite{pospelovreview,Barton,pvcpvhamiltonian1}:
\begin{eqnarray}
{\cal L}_{P\hspace{-.5em}/\, T\hspace{-.5em}/\, } 
&=& 
-\frac{i}{2} \sum_{N=p,n} \bar d_N \bar N \sigma_{\mu \nu} \gamma_5 N F^{\mu \nu}
\nonumber\\
&&
+\sum _{N=p,n} \Biggl[ \sum_{a=1}^3 \bar g_{\pi NN}^{(0)} \bar N \tau^a N \pi^a 
+ \bar g_{\pi NN}^{(1)} \bar N N \pi^0 
\nonumber \\ 
&& \hspace{3em}
+\sum_{a=1}^3 \bar g_{\pi NN}^{(2)} ( \bar N \tau^a N \pi^a - 3\bar N \tau^3 N \pi^0) 
\Biggr]
. \ \ \ \ \ 
\label{eq:pcpvpinnint}
\end{eqnarray}
Here the P, CP-odd coupling constants depend on QCD and elementary level CP violation.
In this work, we consider them as small and given.
There are other CP-odd hadron level effective interactions which contribute in the leading order of chiral perturbation theory, such as the three-pion interaction or the contact CP-odd $NN$ interactions \cite{eft6dim}
\begin{eqnarray}
{\cal L}'_{P\hspace{-.5em}/\, T\hspace{-.5em}/\, } 
&=& 
m_N \Delta_{3\pi} \, \pi^z  \sum_{a=1}^3 \pi_a^2
+
\bar C_1 \bar N N \partial_\mu (\bar N S^\mu N )
\nonumber\\
&&
+ \sum_{a=1}^3 \bar C_2 \bar N \tau_a N \cdot \partial_\mu (\bar N S^\mu \tau_a N )
.
\label{eq:3pi_contact}
\end{eqnarray}
The three-pion interaction (term with $\Delta_{3\pi}$) is isovector, and the radiative correction is known to sizably contribute to $\bar g_{\pi NN}^{(1)}$.
Here the contact interaction [terms with $\bar C_1$ and $\bar C_2$ of Eq. (\ref{eq:3pi_contact})] is isoscalar, and it receives contribution from the $\eta$ meson exchange.
It is interesting , since it is also an important probe of the Weinberg operator \cite{chiral3nucleon}.
In this work, we however do not consider it since its effect suffers from large theoretical uncertainty in the nuclear level calculation \cite{bsaisou,bsaisou2}.

The physical nucleon EDMs $d_n$ and $d_p$ are not only due to the bare terms $\bar d_n$ and $\bar d_p$, but also receive contribution from the isoscalar CP-odd pion-nucleon interaction $\bar g_{\pi NN}^{(0)}$.
In the leading order of chiral perturbation theory, it is given as \cite{crewther}
\begin{eqnarray}
d_N
&=&
\bar d_N 
- \tau_z \frac{e g_A \bar g^{(0)}_{\pi NN} }{4 \pi^2 f_\pi} 
\ln \frac{\Lambda}{m_\pi}
,
\label{eq:d_1}
\end{eqnarray}
where $\Lambda \approx$ 1 GeV is the cutoff of the hadron level effective theory, and $\tau_z = +1$ ($-1$) for the proton (neutron).
Here we neglect the effect of $\bar g_{\pi NN}^{(1)}$ and $\bar g_{\pi NN}^{(2)}$ which contributes at the higher order.

\begin{figure}[htb]
\begin{center}
\includegraphics[width=8cm]{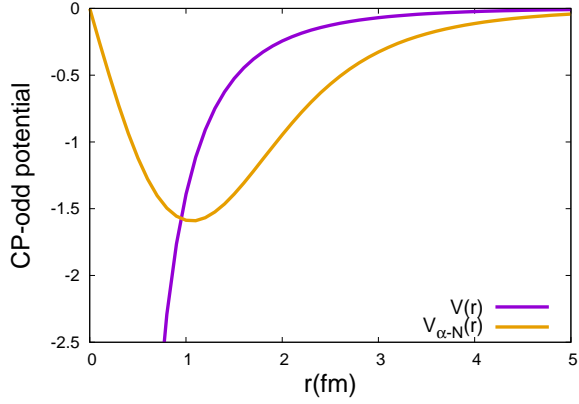}
\caption{\label{fig:folding}
The radial shape of the bare pion exchange CP-odd nuclear force $V(r)$ and its folding potential $V_{\alpha -N}(r)$.
The CP-odd coupling constant was factored out.
}
\end{center}
\end{figure}

We now give the one-pion exchange CP-odd nuclear force.
It has been studied and used in many previous works \cite{korkin,liu,stetcu,song,bsaisou2,yamanakanuclearedm}.
The CP-odd one-pion exchange nuclear force in the coordinate representation is \cite{pvcpvhamiltonian1,pvcpvhamiltonian2,pvcpvhamiltonian3,liu}
\begin{eqnarray}
H_{P\hspace{-.5em}/\, T\hspace{-.5em}/\, }^\pi
& = &
\bigg\{ 
\bar{G}_{\pi}^{(0)}\,{\vc{\tau}}_{1}\cdot {\vc{\tau}}_{2}\, {\vc{\sigma}}_{-}
+\frac{1}{2} \bar{G}_{\pi}^{(1)}\,
( \tau_{+}^{z}\, {\vc{\sigma}}_{-} +\tau_{-}^{z}\,{\vc{\sigma}}_{+} )
\nonumber\\
&&\hspace{1em}
+\bar{G}_{\pi}^{(2)}\, (3\tau_{1}^{z}\tau_{2}^{z}- {\vc{\tau}}_{1}\cdot {\vc{\tau}}_{2})\,{\vc{\sigma}}_{-} 
\bigg\}
\cdot
\hat{ \vc{r}} \,
V(r)
,
\label{eq:CPVhamiltonian}
\end{eqnarray}
where $\hat{\vc{r}} \equiv \frac{\vc{r}_1 - \vc{r}_2}{|\vc{r}_1 - \vc{r}_2|}$ is the unit vector.
The spin and isospin notations are ${\vc{\sigma}}_{-} \equiv {\vc{\sigma}}_1 -{\vc{\sigma}}_2$, ${\vc{\sigma}}_{+} \equiv {\vc{\sigma}}_1 + {\vc{\sigma}}_2$, ${\vc{\tau}}_{-} \equiv {\vc{\tau}}_1 -{\vc{\tau}}_2$, and ${\vc{\tau}}_{+} \equiv {\vc{\tau}}_1 + {\vc{\tau}}_2$.
The dimensionless CP-odd nuclear couplings $\bar G_\pi^{(i)}$ $(i=0,1,2)$ are given by
\begin{eqnarray}
\bar G_\pi^{(0)}
&=&
-\frac{g_A m_N}{f_\pi} \bar{g}_{\pi NN}^{(0)} , \\
\bar G_\pi^{(1)}
&=&
-\frac{g_A m_N}{f_\pi} \bar{g}_{\pi NN}^{(1)} , \\
\bar G_\pi^{(2)}
&=&
\frac{g_A m_N}{f_\pi} \bar{g}_{\pi NN}^{(2)}  ,
\end{eqnarray}
in the leading order of chiral perturbation theory.
The radial shape of the CP-odd $NN$ potential is given by
\begin{equation}
V(r)
= 
-\frac{m_\pi}{8\pi m_N} \frac{e^{-m_\pi r }}{r} \left( 1+ \frac{1}{m_\pi r} \right)
\ ,
\end{equation}
with the pion and nucleon masses $m_\pi = 138$ MeV and $m_N = 939$ MeV, respectively.
The shape of the radial dependence of the CP-odd nuclear force is shown in Fig. \ref{fig:folding}.

To obtain the CP-odd $N - \alpha$ potential, we use the folding by the density function of the $\alpha$ cluster.
It works as~\footnote
{This folding potential has been corrected from that of Ref. \cite{yamanakanuclearedm} by taking into account the directional dependence in the integral. That of Ref. \cite{yamanakanuclearedm} agrees with the present one only in the case where the nucleon and the $\alpha$-cluster are distant from each other. Since $^{13}$C is not dilute, it is adequate to use this corrected folding potential.}
\begin{eqnarray}
V_{\alpha -N } 
( R ) 
\hat{\vc{R}}
&=&
\int d^3 \vc{R}' \,
V 
(|\vc{R } -\vc{R }' | ) \rho_\alpha (R')
\frac{\vc{R } -\vc{R }'}{|\vc{R } -\vc{R }'|}
,
\nonumber\\
\end{eqnarray}
where $\rho_\alpha ( R ) = \frac{4}{ \pi^{3/2} b^3} e^{-R^2/b^2}$ is the density function of the $\alpha$ cluster with the center of mass effect removed, with $b=1.358 \times \sqrt{\frac{4}{3}} \approx 1.57$ fm, and $\vc{R}$ the $N-\alpha$ relative coordinate.
The shape of the CP-odd $\alpha - N$ interaction is shown in Fig. \ref{fig:folding}.
It is important to note that the folding cancels the isoscalar and isotensor CP-odd nuclear force, since the $\alpha$-cluster has closed spin and isospin shells.

The nuclear EDM has two leading sources: 
\begin{itemize}
\item
the intrinsic EDM of the constituent nucleons.

\item
the P, CP-odd $NN$ interactions (CP-odd nuclear force).
\end{itemize}

Let us first see the contribution due to the intrinsic nucleon EDM.
It is given by
\begin{eqnarray}
d_A^{\rm (Nedm)} 
&=&
\sum_{i}^A
d_i \langle \, \Phi_J(A) \, |\, \sigma_{iz} \, |\, \Phi_J(A) \, \rangle
\nonumber\\
&=&
\frac{1}{2} \sum_{i}^A
\Biggl[ \ \ 
d_p \langle \, \Phi_J(A) \, | \, \sigma_{iz} (1+\tau_i^3)\, |\, \Phi_J(A) \, \rangle
\nonumber\\
&& \hspace{3em}
+ d_n \langle \, \Phi_J(A) \, | \, \sigma_{iz} (1-\tau_i^3)\, |\, \Phi_J(A) \, \rangle
\Biggr]
\nonumber\\
&\equiv &
C^{(0)}_A (d_p +d_n) + C^{(1)}_A (d_p-d_n)
,
\label{eq:nedmnuclearedm}
\end{eqnarray}
where $|\, \Phi_J(A) \, \rangle$ is the polarized (in the $z$-direction) nuclear wave function ($A=$ $^{13}$C), and $\tau^3_i$ is the isospin Pauli matrix.
From the above equation, we see that we just have to calculate the spin matrix elements $C^{(0)}_A \equiv \frac{1}{2} \sum_i^A \langle \, \Phi_J(A) \, | \, \sigma_{iz} \, |\, \Phi_J(A) \, \rangle$ and $C^{(1)}_A \equiv \frac{1}{2} \sum_i^A \langle \, \Phi_J(A) \, |\, \sigma_{iz} \tau_i^3 \, |\, \Phi_J(A) \, \rangle$ to obtain the effect of the nucleon EDM on the nuclear EDM.
In this case, we do not need to calculate the mixing between parity-odd and parity-even states.

The nucleon EDM contribution to the nuclear EDM is in the general case not enhanced.
If the spins of several nucleons are aligned inside the nucleus, the nuclear spin matrix elements may be enhanced.
This is, however, not the case for nuclear ground states, due to the strong pairing correlation.
The EDM of composite systems may also be enhanced through the polarization of the whole system by the EDM of the components.
In heavy atoms, the polarization due to the electron EDM is known to be enhanced by the relativistic effect \cite{sandars1,sandars2,flambaumenhancement}.
In nuclear systems, such enhancement is absent, since the system is nonrelativistic.
Rather, it was recently suggested that the polarization of the nucleus due to the interaction between the EDM of the nucleon and the nuclear internal electric field can suppress the total intrinsic nucleon EDM contribution \cite{inoue}.
This phenomenon resembles the screening of the EDM of constituents in an electrically bound neutral system, first pointed out by Schiff \cite{schiff}.
In this work, we do not consider this effect.
The nuclear spin matrix elements are also suppressed by the mixing of different angular momentum configurations.
In the general case, the nuclear wave function shares some portion of states which have different orbital angular momentum, so $C^{(0)}_A$ and $C^{(1)}_A$ are suppressed.
The coefficient relating the nucleon EDM to the nuclear EDM $\frac{1}{2} ( C^{(0)}_A  \pm C^{(1)}_A ) $ is therefore at most one for the majority of nuclei.

The evaluation of the nuclear polarization due to the P, CP-odd nuclear force is more complicated than the previous case, since it arises from the mixing between the parity-even and parity-odd states.
In this work we assume that a P, CP-odd nuclear force with a small coupling constant exists.
As this P, CP-odd coupling is small, the nuclear EDM should have a linear dependence on it.
The polarization contribution of the P, CP-odd nuclear force to the nuclear EDM is given by
\begin{eqnarray}
d_{A}^{\rm (pol)} 
&=&
\sum_{i=1}^{A} \frac{e}{2} 
\langle \, \tilde \Phi_{J=1/2} \, |\, (1+\tau_i^3 ) \, r_{iz} \, | \, \tilde \Phi_{J=1/2} \, \rangle
\nonumber\\
&\approx &
\frac{e}{2} \sum_{i=1}^{A} \sum_{n\neq 0 } \frac{1}{E_0 - E_n } 
\nonumber\\
&& \hspace{2em} \times
\langle \, \Phi_{J^P =1/2^-_1} \, | (1+\tau_i^3 ) \, r_{iz} | \, \Phi_{J^P = 1/2^+_n}\, \rangle 
\nonumber\\
&& \hspace{2em} \times
\langle \, \Phi_{J^P = 1/2^+_n} \, |\, H_{P\hspace{-.5em}/\, T\hspace{-.5em}/\, } \, | \, \Phi_{J^P = 1/2^-_1} \, \rangle 
\nonumber\\
&&
+({\rm c.c.})
,
\label{eq:polarizationedm}
\end{eqnarray}
where $|\, \tilde \Phi_{J=1/2} \, \rangle$ is the polarized (in the $z$-axis) nuclear wave function obtained by diagonalizing the hamiltonian $ \mathcal{H}+H_{P\hspace{-.5em}/\, T\hspace{-.5em}/\, } $.
$r_{iz}$ is (the $z$-component of) the position of the constituent nucleon in the nuclear center of mass frame.
The second equality is the 1st order perturbation in the P, CP-odd nuclear force $H_{P\hspace{-.5em}/\, T\hspace{-.5em}/\, }$, where $|\, \Phi_{J^P =1/2^-_1} \, \rangle$ is the (polarized) nuclear wave function without opposite parity states, $|\, \Phi_{J^P = 1/2^+_n} \, \rangle$ the (polarized) opposite parity states, and $E_n$ their corresponding energy.

\begin{figure}[t]
\begin{center}
\includegraphics[width=5cm]{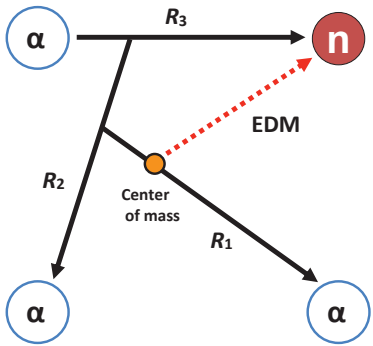}
\caption{\label{fig:edm_jacobi}
Jacobi coordinates to express the EDM of $^{13}$C.
}
\end{center}
\end{figure}

The polarization operator of $^{13}$C in the $\alpha$-cluster model is given by
\begin{eqnarray}
\sum_{i=1}^4 Q_i e \vc{r}_i
&=&
2 e \vc{r}_1
+2 e \vc{r}_2
+2 e \vc{r}_3
+\frac{e}{2} (1+\tau_4^z) \vc{r}_4
\nonumber\\
&=&
\frac{e}{2} \tau_4^z \vc{r}_4
\nonumber\\
&=&
e \tau_4^z
\left[
-\frac{2}{13} \vc{R}_1
-\frac{2}{9} \vc{R}_2
+\frac{2}{5} \vc{R}_3
\right]
,
\label{eq:t1r1+t2r2}
\end{eqnarray}
where $\vc{r}_i \ (i=1,2,3)$ denotes the coordinates of the $\alpha$ clusters, and $\vc{r}_4$ that of the nucleon, in the center of mass frame ($4 \vc{r}_1 + 4 \vc{r}_2 + 4\vc{r}_3 + \vc{r}_4=0$).
The last line is expressed in terms of the Jacobi coordinate (see Fig. \ref{fig:edm_jacobi}).
In this work, we do not consider the effect of meson exchange current \cite{pastore1,pastore2,pastore3,pastore4}.

\section{Result and analysis}

We first show the result of the evaluation of the intrinsic nucleon EDM contribution to the EDM of $^{13}$C.
After calculation, we obtain the following value:
\begin{eqnarray}
d_{^{13}{\rm C}}^{\rm (Nedm)} 
&=&
-0.33 \, d_n
.
\label{eq:neutron13cedm}
\end{eqnarray}
This result agrees with the formula of the EDM of $^{13}$C when we assume that the nucleus is a $p$-wave core+valence nucleon system \cite{sushkov}
\begin{eqnarray}
d_{^{13}{\rm C}}^{\rm (Nedm)} 
&=&
-\frac{1}{3} d_n
.
\end{eqnarray}
Here we repeat that the effect of the interaction between the nucleon EDM and the nuclear internal electric field \cite{inoue} was not taken into account.
This may suppress the nucleon EDM contribution to the nuclear EDM, although we do not evaluate it.

Let us try to see the consistency with the experimental data of the nuclear magnetic moment.
If we assume the $p$-wave core+valence nucleon system with the valence neutron magnetic moment $\mu_n = -1.9130$ is giving the entire contribution, then we have
\begin{eqnarray}
\mu_{^{13}{\rm C}}
&=&
2 \mu_n C^{(0)}_{^{13}{\rm C}}
=
0.63
,
\end{eqnarray}
where $C^{(0)}_{^{13}{\rm C}}$ is defined  in Eq. (\ref{eq:nedmnuclearedm}).
This estimation is in reasonable agreement with the experimental value $\mu_{^{13}{\rm C}} = 0.702411(1)$.
Our result is therefore consistent with the experimental data of the nuclear magnetic moment.

\begin{figure}[htb]
\begin{center}
\includegraphics[width=8cm]{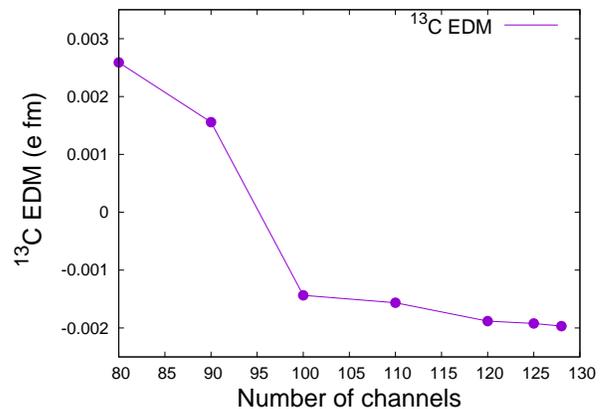}
\caption{\label{fig:convergence_13c_edm_gs}
Convergence of the EDM of $^{13}$C displayed in the function of the number of channels, with 343 bases for each.
The CP-odd coupling constant $\bar G_\pi^{(1)}$ was factored out.
}
\end{center}
\end{figure}

Next, we show the result of the calculation of the polarization contribution.
The effect of the CP-odd nuclear force to the nuclear EDM of $^{13}$C is
\begin{eqnarray}
d_{^{13}{\rm C}}^{\rm (pol)} 
&=&
-0.0020 \, \bar G_\pi^{(1)} e\, {\rm fm}
.
\label{eq:13cedmresult}
\end{eqnarray}
Note that only the isovector CP-odd nuclear force contributes to the $^{13}$C EDM in the $\alpha$-cluster model.
We used $7^3 \times 128 = 43904$ bases to converge the EDM of $^{13}$C (see Section \ref{sec:structure}).
The convergence of the nuclear EDM in the function of the angular momentum channels is shown in Fig. \ref{fig:convergence_13c_edm_gs}.
We see that the EDM of $^{13}$C is smaller than the result of the calculation for lighter nuclei in Ref. \cite{yamanakanuclearedm} [e.g., the deuteron EDM $d_{^{2}{\rm H}}^{\rm (pol)} = 0.0145 \, \bar G_\pi^{(1)} e\, {\rm fm}$].
This fact suggests that some suppression mechanisms of the EDM are relevant for $^{13}$C.
We note that, although being smaller than other lighter nuclei, the EDM of $^{13}$C is only smaller than them by an order of magnitude.
It is much larger than the nuclear EDM (not atomic EDM!) of $^{129}$Xe by several orders \cite{yoshinaganuclearedm}.

Let us try to analyze the physical mechanism which suppresses the EDM following Eq. (\ref{eq:polarizationedm}).
The first point is the distance between the valence nucleon and the core.
The expectation value of the EDM becomes larger when the $1/2^-_1$ and the opposite parity $1/2^+_1$ states have better overlap.
Here good overlap points to the bra and ket wave functions which make large matrix elements of the CP-odd nuclear force and the EDM operator.
Such a combination of wave functions must therefore be well related with a $|\Delta L| = 1$ transition, with the constituents not distant from each other, since the CP-odd nuclear force is exponentially damping at long distance.

In looking at the $1/2^-_1$ state, the root mean square radius of the core-valence distance of $^{13}$C is $\langle \sqrt{r^2} \rangle_{\rm 1/2^-_1} = 2.81$ fm, whereas that of the $1/2^+_1$ state is $\langle \sqrt{r^2} \rangle_{\rm 1/2^+_1} = 3.95$ fm \cite{yamada}.
We emphasize that in this work we refer the ground and first excited states as the lowest and the next lowest energy states, respectively, obtained after diagonalizing the hamiltonian including the CP-odd nuclear force for the system with angular momentum 1/2, while the $1/2^-_1$ and $1/2^+_1$ states correspond to those unperturbed by the CP-odd nuclear force, without parity mixing.
We see that $^{13}$C has a shell structure in the $1/2^-_1$ state, but a neutron halo structure in the $1/2^+_1$ excited state.
The bad overlap of the core-valence wave function may therefore suppress the EDM.

Another bad overlap between the ground and $1/2^+_1$ states also exists.
The core of $^{13}$C, i.e., the $^{12}$C subsystem, actually has a different structure between the two states.
In Ref. \cite{yamada}, the spectroscopic factors for both states were calculated.
There it was found that the core is dominated by the $2^+$ state in the ground state.
This is due to the strong $LS$ attraction between the valence nucleon and the $\alpha$ cluster which forces the alignment of their orbital angular momentum ($p$-wave) and the nucleon spin to form an angular momentum 3/2 system.
This fact, consequently, requires the core to have at least angular momentum 2 to be consistent with the total angular momentum 1/2 of the $^{13}$C nucleus.
However, in the $1/2^+_1$ state, the leading contribution is given by a $0^+$ state.
This makes a bad overlap between the cores of the $1/2^-_1$ and $1/2^+_1$ states and, consequently, suppresses the EDM.

\begin{figure*}[htb]
\begin{center}
\includegraphics[width=12cm]{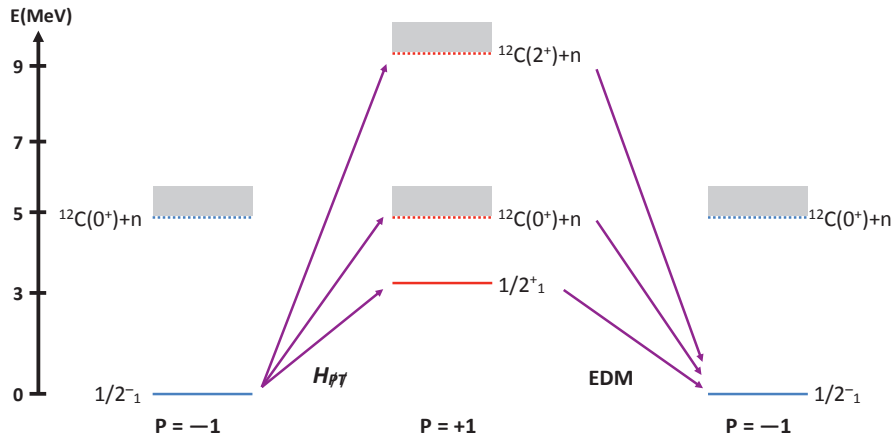}
\caption{\label{fig:energy_level_transition}
Parity-odd transitions contributing to the EDM of $^{13}$C in the leading order of perturbation.
The dashed lines represent the dissociation thresholds into $^{12}$C($0^+$) + $n$ or $^{12}$C($2^+$) + $n$ states.
The gray bands correspond to the continuum states.
}
\end{center}
\end{figure*}

The spectrum of $^{13}$C (not perturbed with CP-odd nuclear force) shows a $1/2^+_1$ bound state at 3.1 MeV.
In the study of the EDM of light nuclei, it is the first case where we can find the ground and first excited states both as bound states.
In the leading order of perturbation (which works well in the present case), the EDM receives a contribution from the transition between the $1/2^-_1$ and $1/2^+_1$ states and that between the $1/2^-_1$ state and the $1/2$ continuum state [see Eq. (\ref{eq:polarizationedm}) and Fig. \ref{fig:energy_level_transition}].
It is interesting to compare those two effects.
The EDM of the ground state (in the spectrum with the CP-odd nuclear force) of $^{13}$C, generated by the transition between $1/2^-_1$ and $1/2^+_1$ states (solved without CP-odd nuclear force), is
\begin{eqnarray}
d^{\rm bound}_{^{13}{\rm C} }
&\approx &
\sum_i 
\frac{\langle \Phi_{J^P =1/2^-_1} | Q_i e r_i | \Phi_{J^P =1/2^+_1} \rangle}{E_{1/2^-} -E_{1/2^+}}
\nonumber\\
&&\hspace{2em} \times 
\langle \Phi_{J^P =1/2^+_1} | H_{P\hspace{-.5em}/\, T\hspace{-.5em}/\, }^\pi | \Phi_{J^P =1/2^-_1} \rangle 
.
\label{eq:boundstatetransitionedm}
\end{eqnarray}
Our calculation gives
\begin{eqnarray}
d_{^{13}{\rm C}}^{\rm bound}
&=&
0.00025 \, \bar G_\pi^{(1)} e\, {\rm fm}
.
\label{eq:edm_bound}
\end{eqnarray}
The remaining part is the contribution from the transition to the continuum states and resonances, which is given by
\begin{eqnarray}
d_{^{13}{\rm C}}^{\rm rem}
=
d_{^{13}{\rm C}}^{\rm (pol)}
-
d_{^{13}{\rm C}}^{\rm bound}
&=&
-0.0022 \, \bar G_\pi^{(1)} e\, {\rm fm}
.
\end{eqnarray}
In Ref. \cite{yamada}, five resonances have been identified for each parity.
By calculating their contribution as in Eq. (\ref{eq:boundstatetransitionedm}), it is found that the effect of resonances is not larger than 10\% of $d_{^{13}{\rm C}}^{\rm rem}$.
The transition to continuum is therefore dominant in the EDM of $^{13}$C.

How about the excited state (in the spectrum with the CP-odd nuclear force)?
In our framework, the EDM of the first excited state (in the spectrum with the CP-odd nuclear force) can also be calculated.
The EDM of the excited states is given by 
\begin{eqnarray}
d_{^{13}{\rm C}^*}^{\rm (pol)}
&=&
0.0244 \, \bar G_\pi^{(1)} e\, {\rm fm}
.
\end{eqnarray}
We see that the EDM of the excited states is much larger in magnitude than the ground state EDM (by about ten times).
Among them, the contribution from the bound state transition is just
\begin{eqnarray}
d_{^{13}{\rm C}^*}^{\rm bound}
&=&
-d_{^{13}{\rm C}}^{\rm bound}
=
-0.00025 \, \bar G_\pi^{(1)} e\, {\rm fm}
.
\end{eqnarray}
For the remaining contribution, we, however, have
\begin{eqnarray}
d_{^{13}{\rm C}^*}^{\rm rem}
=
d_{^{13}{\rm C}^*}^{\rm (pol)}
-
d_{^{13}{\rm C}^*}^{\rm bound}
&=&
0.0247 \, \bar G_\pi^{(1)} e\, {\rm fm}
.
\end{eqnarray}
which is dominantly made by the transition to continuum.
The contribution from resonances is less than 3\%.
We see that $d_{^{13}{\rm C}^*}^{\rm rem}$ is much larger than that of the ground state.
The continuum contribution represents the majority for the EDM of excited $^{13}$C.
These large values can be explained as follows.
The analysis of the spectroscopic factor in Ref. \cite{yamada} shows that the excited state ($1/2^+_1$ in the absence of CP-odd nuclear force) has a dominant $0^+$ state contribution for the core, $^{12}$C($0^+$) \cite{yamada}.
The (bound) excited state is dominantly made of a halo-like configuration of valence nucleon ($s$-wave) and the $^{12}$C($0^+$)  core.
This state should have a large overlap with the $^{12}$C($0^+$) core+$n$ ($p$-wave) continuum state, for which the threshold opens just 1.8 MeV (for $^{13}$C) above the energy level of the $1/2^+_1$ state (4.9 MeV above the $1/2^-_1$ state of $^{13}$C, see Fig. \ref{fig:energy_level_transition}).

For the case of the ground state (with CP-odd nuclear force), the core has a dominant $2^+$ configuration, $^{12}$C($2^+$).
Therefore the EDM does not receive a large contribution from the continuum $^{12}$C($0^+$) + $n$ ($s$-wave) due to the small overlap with this continuum.
The closest threshold of continuum states that dominantly couple to the ground state, $^{12}$C($2^+$) + $n$, is about 4 MeV above the $^{12}$C($0^+$) + $n$ threshold (about 9 MeV above the $1/2^-_1$ state), since we need this energy to excite $^{12}$C($0^+$) to the $2^+_1$ state (see Fig. \ref{fig:energy_level_transition}).
From the expression of the perturbation (\ref{eq:polarizationedm}), the EDM is inversely proportional to the energy difference.
The EDM of the ground state, although having a good overlap with the $^{12}$C($2^+$) + $n$ continuum, is suppressed by the large energy difference 9 MeV, and becomes smaller than the EDM of the exited state, for which the energy difference is only 1.8 MeV.

Our analysis of the EDM of $^{13}$C has qualitatively explained the suppression and the enhancement of the ground and first excited state EDMs, respectively, by inspecting the energy levels and the overlap of opposite parity states.
This fact suggests that the nuclear EDM is very sensitive to the nuclear structure.
As the structure of nuclei differs much between nuclei, its investigation also guides us in finding nuclei with a large EDM.
As a nucleus which has a continuum state threshold near the ground state, we have the $^7$Li nucleus, which has a $^3$H + $\alpha$ threshold at 2.47 MeV above the ground state.
The $^{19}$F nucleus is also a very good candidate, as it has an opposite parity first excited state at 110 keV above the ground state.

We also mention the theoretical uncertainty of our result.
The intrinsic nucleon EDM contribution to the EDM of $^{13}$C is well explained in the valence nucleon picture, and the discrepancy of the calculated magnetic moment with the experimental data is about 10\%.
The error bar of our result [Eq. (\ref{eq:neutron13cedm})] should be of the same order, like 10\%.

The theoretical uncertainty due to the isovector CP-odd nuclear force [term with $\bar G_\pi^{(1)}$, Eq. (\ref{eq:CPVhamiltonian})] is estimated from the energy spectrum calculated in detail in our previous work \cite{yamada}.
The energy levels are well reproduced, within about 30\% for those above the $^{12}$C($0^+$) + $n$ threshold.
As we showed in this section that the nuclear EDM does not receives much contribution from resonances, it is expected that it is less affected by the discrepancy of the energy level scheme.

We also mention the theoretical uncertainty related to the $E1$ transition.
The observed energy spectrum of $^{13}$C contains a giant dipole resonance near 25 MeV which might sizably contribute to the EDM.
The cluster model cannot fully describe it, so we have to consider this contribution as a systematic error.
Here we shall show that it is not large.
Since the giant dipole resonance is located at 25 MeV, its contribution to the EDM is damped by the large energy in the denominator [see Eq. (\ref{eq:polarizationedm})].
We also see from Eq. (\ref{eq:polarizationedm}) that the EDM is generated not only by the isovector dipole operator, but also by the isovector spin-flip operator of the CP-odd nuclear force [see Eq. (\ref{eq:CPVhamiltonian})].
This latter induces an isovector spin-flip giant dipole resonance, which requires higher excitation energy than the giant dipole resonance.
Unless those giant resonances are strongly coupled, they should not give important effects. 

The $^{12}$C + $n$ continuum states therefore play the most important role in the EDM of $^{13}$C. 
The $3\alpha$ + $n$ OCM (orthogonality condition model), which is a semi-microscopic cluster model, can simultaneously describe the $^{12}$C + $n$ asymptotic behavior, as well as the $^{12}$C + $n$ threshold energy, and the ground state of $^{13}$C with a shell-model like structure, whereas the ordinary shell model or other mean-field approaches cannot.
There the $^{12}$C + $n$ continuum states should well be described, since the structure of $^{12}$C is accurately given in the $3\alpha$ OCM \cite{yamadaschuck,kurokawa,ohtsubo}.
Here we conservatively estimate the theoretical error bar as 50\%, since the EDM of $^{13}$C is generated through a suppression mechanism.

\section{Role of $^{13}$C EDM in the determination of hadron level CP violation\label{sec:roles}}

Let us inspect the necessity to measure the EDM of $^{13}$C.
Its dependence on the bare CP-odd couplings of Eq. (\ref{eq:pcpvpinnint}), in the leading order of chiral perturbation theory, is given by
\begin{equation}
d_{^{13}{\rm C}}
=
-0.33 \bar d_n
+(
-0.045 \bar g_{\pi NN}^{(0)}
-0.025 \bar g_{\pi NN}^{(1)}
)
e\, {\rm fm}
.
\end{equation}
Here we neglect the isotensor CP-odd pion-nucleon interaction [term with $\bar g_{\pi NN}^{(2)}$ in Eq. (\ref{eq:pcpvpinnint})].
The dependence of $d_{^{13}{\rm C}}$ on $\bar g_{\pi NN}^{(0)}$ dominantly comes from the valence neutron EDM.

By using the relation between the neutron EDM and $\bar \theta$, obtained in the chiral effective field theory combined with lattice QCD data \cite{devries3},
the $\theta$-term contribution to $^{13}$C EDM is given by
\begin{equation}
d_{^{13}{\rm C}}
=
(9 \pm 5) \times 10^{-17}
\bar \theta \, e \, {\rm cm}
.
\end{equation}
This contribution is dominantly due to the valence neutron EDM.
The dependence of the EDM of $^{13}$C on $\bar g_{\pi NN}^{(0)}$ as well as on the $\theta$-term is more pronounced than that of the deuteron and $^6$Li, since the effect of $\bar g_{\pi NN}^{(0)}$ cancels in the leading order.
We see below that even if the sensitivity of $^{13}$C on individual CP-odd interactions is smaller than other light nuclei (see Ref. \cite{yamanakanuclearedm}), it has its own speciality. 

It is important to note that the experimental data of the EDM of a single system, even if a finite value is measured, are not sufficient to determine the new physics beyond the standard model.
Obviously, the combination of the observations for several systems with linearly independent coefficients will absolutely be required to constrain multiple CP-odd couplings.
Explicitly, there are four unknown CP-odd couplings [$\bar d_0$, $\bar d_1$, $\bar g_{\pi NN}^{(0)}$, $\bar g_{\pi NN}^{(1)}$] in the hadron level effective interaction we are considering [see Eq. (\ref{eq:pcpvpinnint})].
To fix them, at least four experimental measurements of EDM, using systems with linearly independent dependence on them, are required.
We emphasize that those four unknown parameters do not form the complete set of leading CP-odd couplings, and additional CP-odd interactions such as the three-pion interaction or the contact interactions [see Eq. (\ref{eq:3pi_contact})] are known to contribute in the leading order of chiral perturbation theory \cite{engel,bsaisou,bsaisou2,devries3,eft6dim}.
To fix them, additional EDM experimental data are required.
The experimental measurements of the EDM of lightest systems, namely the neutron, the proton, the deuteron, and $^3$He, are not sufficient to determine the new physics.
In this circumstance, the experimental constraint from the EDM of $^{13}$C will play an important role.

As other potential candidates of nuclei to be measured, we have $^6$Li and $^9$Be \cite{yamanakanuclearedm}.
The $^{13}$C and $^9$Be nuclei have similar dependence on the CP-odd couplings, since they probe $d_n$ through the valence neutron, and $\bar g_{\pi NN}^{(1)}$ via the polarization.
It is maybe useful to note that the experimental manipulation of $^9$Be may have some disadvantages comparing with $^{13}$C, since the $^9$Be atom is unstable, and its raw component is toxic.
The $^{13}$C nucleus is therefore a good candidate to probe the linear combination of $d_n$ and $\bar g_{\pi NN}^{(1)}$.

In the analysis of new physics beyond the standard model, there is always some region of the parameter space which escapes the EDM-constraints \cite{ibrahim,ramseyli,ellisgeometric,ellisgeometric2,rpvlinearprogramming}. 
The measurement of $^{13}$C EDM may therefore be an important piece to determine or discriminate candidates of new physics.
Moreover, the coefficient of the isovector nuclear force [$\bar g_{\pi NN}^{(1)}$] and that for the intrinsic nucleon EDM are negative.
Other light nuclei studied so far all have corresponding coefficients positive.
The $^{13}$C nucleus has thus the possibility to provide a good check of the relative sign.

We should not forget to show the standard model contribution generated by the CP phase of the Cabibbo-Kobayashi-Maskawa matrix \cite{yamanakasmedm}:
\begin{eqnarray}
d_{^{13}{\rm C}}^{\rm (SM)}
&=&
-3.3 \times 10^{-32} e \, {\rm cm}
.
\end{eqnarray}
We see that the effect of the standard model is small.
Here we have only considered the polarization effect from the pion exchange CP-odd nuclear force.
The nucleon EDM also contributes to the nuclear EDM \cite{mannel,seng}, but its contribution should be small.
The reason is the same as that for the suppression of the effect of new physics candidates which contribute through the quark EDM.

\section{Prospects to the discovery of new physics\label{sec:newphys}}

Let us now see the prospect for the discovery of new physics BSM.
The first typical CP violating model to be inspected is the supersymmetric model \cite{pospelovreview,ellismssmedm,ibrahim,chang,demir,mssmreloaded,ramseyli,ellisgeometric,ellisgeometric2,mssmrainbow}.
If the EDM of $^{13}$C can be measured at the level of $O(10^{-29})e $ cm \cite{bnl}, then the mass scale of the supersymmetry breaking can be probed at the TeV scale (see Ref. \cite{yamanakanuclearedm}). 

The class of models which generate the four-quark interaction or the Barr-Zee type diagram, contributes through the combination $\frac{{\rm Im} (gg')}{m_{\rm NP}^2}$.
If the nuclear EDM is measured with the prospective sensitivity $O(10^{-29})e $ cm, then the latter combination of couplings and new particle mass can be probed at the level of $O(0.1 \, {\rm PeV})$.
Those new physics include the left-right symmetric model \cite{dekens}, the Higgs doublet models \cite{barr-zee,jung2hdmedm,abe}, supersymmetric models with $R$-parity violation \cite{rpv1,rpv2,rpv3,rpv4,rpvlinearprogramming}, etc.

Models which contribute to the nuclear EDM only through the quark EDM are much less sensitive than those which generate the quark chromo-EDM.
This fact is due to several physical reasons.
The first reason is because the effect of the quark EDM to the nucleon EDM is suppressed by the nucleon tensor charge.
The nucleon tensor charge is the linear coefficient which relates the quark EDM to the nucleon EDM, and the extraction from experimental data shows values smaller than one \cite{bacchetta,anselmino,courtoy,kang}.
Recent lattice QCD analyses give also consistent data \cite{saoki,green,yaoki,etm2,etm3,bhattacharya1,bhattacharya2,bhattacharya3,jlqcd4}.
It is suggested that nucleon charges which give the quark spin in the nonrelativistic limit, such as the axial or tensor charges, are suppressed by the dynamical gluon dressing \cite{yamanakasde1,yamanakasde2,pitschmann}.
The second reason is the suppression of the Wilson coefficient of the quark EDM operator by the renormalization group evolution.
By running this Wilson coefficient from the scale $\mu = 1$ TeV to the hadronic scale $\mu = 1$ GeV, it becomes less than 80\% \cite{tensorrenormalization,degrassi,yang}.
The third one is the suppression of the nucleon EDM contribution to the EDM of $^{13}$C by a factor of $-\frac{1}{3}$, due to the antiparallel valence nucleon spin and the core-valence orbital angular momentum.
There may also be additional screening effect to the nucleon EDM, generated by the interaction between the nucleon EDM and the nuclear internal electric field \cite{inoue}. 
For the case of the Barr-Zee type diagram, the fermion EDM may also be suppressed by the electromagnetic coupling and the fractional charge of the quarks, relative to the strong coupling.
For models which generate the quark electromagnetic EDM at leading order, it is therefore strongly recommended to probe them through the neutron EDM.

\section{Summary}

In this work, we have calculated the EDM of $^{13}$C using the Gaussian expansion method.
There we have assumed the $\alpha$-cluster model.
As a result, we obtained that $^{13}$C has smaller enhancement factors than the deuteron, $^3$He, $^3$H, $^6$Li, and $^9$Be.
The sensitivity of the EDM of $^{13}$C is nevertheless much larger than the EDM of the bare $^{129}$Xe nucleus, and this fact shows us that the EDM of light nuclei are promising observables.
Its experimental measurement is absolutely required for determining multiple hadron level CP violating couplings.

By analyzing the suppression of the EDM of $^{13}$C, we have found a mechanism to enlarge the matrix elements of CP-odd operators between opposite parity states, required to enhance the nuclear EDM.
In the case of $^{13}$C, the bad overlap between the $^{12}$C core of the $1/2_1^-$ and the $1/2_1^+$ or the closest continuum states suppresses the EDM.
This shows that the knowledge of the structure of nuclei is essential in giving the linear coefficient between the nuclear EDM and the CP-odd nuclear couplings.

The result of this work provides a very good guide to find nuclei with large EDM.
In particular, we expect a large EDM for $^7$Li and $^{19}$F.
This enhancement can potentially reach more than $O(10)$ times than those so far known.
The study of those nuclei will be the subject of future works.

Moreover, the determination of the new physics beyond the standard model cannot be achieved by only measuring the EDM of one sensitive system, and combining experimental data is essential.
In particular, there are more than four types of CP-odd effective interactions contributing in the leading order of chiral perturbation theory, and the measurements of the EDM of light systems (neutron, proton, deuteron, and $^3$He) are not sufficient to determine the hadron level CP violation.
The nuclear EDM of $^{13}$C can therefore play a significant role in driving the CP violation of new physics into a corner or discriminate among candidates.

\begin{acknowledgments}
This work is supported by the RIKEN iTHES Project and JSPS Postdoctoral Fellowships for Research Abroad.
It is also supported by the HPCI Project.
\end{acknowledgments}


\begin{thebibliography}{99}

\bibitem{sakharov}
A. D. Sakharov, 
%\emph{Violation of CP Invariance, c Asymmetry, and Baryon Asymmetry of the Universe}, 
Pisma Zh. Eksp. Teor. Fiz. {\bf 5}, 32 (1967) [JETP Lett. {\bf 5}, 24 (1967)].

\bibitem{shaposhnikov1}
M. E. Shaposhnikov, 
%\emph{Possible Appearance of the Baryon Asymmetry of the Universe in an Electroweak Theory}, 
Pisma Zh. Eksp. Teor. Fiz. {\bf 44}, 364 (1986) [JETP Lett. {\bf 44}, 465 (1986)].

\bibitem{shaposhnikov2}
M. E. Shaposhnikov, 
%\emph{Baryon asymmetry of the universe in standard electroweak theory}, 
Nucl. Phys. B {\bf 287}, 757 (1987).

\bibitem{farrar}
G. R. Farrar and M. E. Shaposhnikov, Phys. Rev. D {\bf 50}, 774 (1994).

\bibitem{huet}
P. Huet and E. Sather, Phys. Rev. D {\bf 51}, 379 (1995).

\bibitem{hereview}
X.-G. He, B. H. J. McKellar, and S. Pakvasa, 
%\emph{The neutron electric dipole moment}, 
Int. J. Mod. Phys. A {\bf 4}, 5011 (1989) [Erratum ibid. A {\bf 6}, 1063 (1991)].

\bibitem{Bernreuther}
W. Bernreuther and M. Suzuki, 
%\emph{The electric dipole moment of the electron}, 
Rev. Mod. Phys. {\bf 63}, 313 (1991) [Erratum ibid. {\bf 64}, 633 (1992)].

\bibitem{khriplovichbook}
I. B. Khriplovich and S. K. Lamoreaux, 
\emph{CP Violation without Strangeness} (Springer, Berlin, 1997).

\bibitem{ginges}
J. S. M. Ginges and V. V. Flambaum, 
%\emph{Violations of fundamental symmetries in atoms and tests of unification theories of elementary particles}, 
Phys. Rep. {\bf 397}, 63 (2004).% [physics/0309054].

\bibitem{pospelovreview}
M. Pospelov and A. Ritz, 
%\emph{Electric dipole moments as probes of new physics}, 
Annals Phys. {\bf 318}, 119 (2005).% [hep-ph/0504231].

\bibitem{fukuyama}
T. Fukuyama, 
%\emph{Searching for new physics beyond the standard model in electric dipole moment}, 
Int. J. Mod. Phys. A {\bf 27}, 1230015 (2012).% [arXiv:1201.4252].

\bibitem{naviliat}
O. Naviliat-Cuncic and R. G. E. Timmermans, 
%\emph{Electric dipole moments: flavor-diagonal CP violation}, 
Comptes Rendus Physique {\bf 13}, 168 (2012).

\bibitem{engel}
J. Engel, M. J. Ramsey-Musolf, and U. van Kolck, 
%\emph{Electric dipole moments of nucleons, nuclei and atoms: the standard model and beyond}, 
Prog. Part. Nucl. Phys. {\bf 71}, 21 (2013).% [arXiv:1303.2371].

\bibitem{yamanakabook}
N. Yamanaka, 
\emph{Analysis of the Electric Dipole Moment in the R-Parity Violating Supersymmetric Standard Model} (Springer, Berlin, 2014).

\bibitem{hewett}
J. L. Hewett {\it et al.}, 
%\emph{Fundamental Physics at the Intensity Frontier}, 
arXiv:1205.2671 [hep-ex].

\bibitem{roberts}
B. M. Roberts, V. A. Dzuba, and V. V. Flambaum, 
%\emph{Parity and Time-Reversal Violation in Atomic Systems}, 
Ann. Rev. Nucl. Part. Sci. {\bf 65}, 63 (2015).% [arXiv:1412.6644].

\bibitem{baker}
C. A. Baker {\it et al}., 
%\emph{An improved experimental limit on the electric dipole moment of the neutron}, 
Phys. Rev. Lett. {\bf 97}, 131801 (2006).% [hep-ex/0602020].

\bibitem{rosenberry}
M. A. Rosenberry and T. E. Chupp, 
%\emph{Atomic electric dipole moment measurement using spin exchange pumped masers of $^{129}$Xe and $^3$He}, 
Phys. Rev. Lett. {\bf 86}, 22 (2001).

\bibitem{regan}
B.C. Regan, E.D. Commins, C.J. Schmidt, and D. DeMille, 
%\emph{New limit on the electron electric dipole moment}, 
Phys. Rev. Lett. {\bf 88}, 071805 (2002).

\bibitem{griffith}
W. C. Griffith {\it et al}., 
%\emph{Improved limit on the permanent electric dipole moment of $^{199}$Hg}, 
Phys. Rev. Lett. {\bf 102}, 101601 (2009).

\bibitem{graner}
B. Graner, Y. Chen, E. G. Lindahl, and B. R. Heckel, Phys. Rev. Lett. {\bf 116}, 161601 (2016). %arXiv:1601.04339 [physics.atom-ph].

\bibitem{parker}
R. H. Parker {\it et al.}, 
%\emph{First Measurement of the Atomic Electric Dipole Moment of $^{225}$Ra}, 
Phys. Rev. Lett. {\bf 114}, 233002 (2015).% [arXiv:1504.07477].

\bibitem{hudson}
J. J. Hudson {\it et al.}, 
%\emph{Improved measurement of the shape of the electron}, 
Nature {\bf 473}, 493 (2011).

\bibitem{acme}
J. Baron {\it et al}. (ACME Collaboration), 
%\emph{Order of magnitude smaller limit on the electric dipole moment of the electron}, 
Science {\bf 343}, 269 (2014).% [arXiv:1310.7534].

\bibitem{muong-2}
G. W. Bennett {\it et al}. [Muon (g-2) Collaboration], 
%\emph{An improved limit on the muon electric dipole moment}, 
Phys. Rev. D {\bf 80}, 052008 (2009).% [arXiv:0811.1207].

\bibitem{storage1}
I. B. Khriplovich, 
%\emph{Feasibility of search for nuclear electric dipole moments at ion storage rings}, 
Phys. Lett. B {\bf 444}, 98 (1998).% [hep-ph/9809336].

\bibitem{storage2}
F. J. M. Farley {\it et al}., 
%\emph{A new method of measuring electric dipole moments in storage rings},
Phys. Rev. Lett. {\bf 93}, 052001 (2004).% [hep-ex/0307006].

\bibitem{storage3}
Y. K. Semertzidis {\it et al}. (EDM Collaboration), 
%\emph{A new method for a sensitive deuteron EDM experiment}, 
AIP Conf. Proc. {\bf 698}, 200 (2004).% [hep-ex/0308063].

\bibitem{storage4}
Y. F. Orlov, W. M. Morse, and Y. K. Semertzidis, 
%\emph{Resonance method of electric-dipole-moment measurements in storage rings}, 
Phys. Rev. Lett. {\bf 96}, 214802 (2006).% [hep-ex/0605022].

\bibitem{storage5}
T. Fukuyama and A. J. Silenko, 
%\emph{Derivation of Generalized Thomas-Bargmann-Michel-Telegdi Equation for a Particle with Electric Dipole Moment}, 
Int. J. Mod. Phys. A {\bf 28}, 1350147 (2013).% [arXiv:1308.1580].

\bibitem{storage6}
V. Anastassopoulos {\it et al.}, 
%\emph{A Storage Ring Experiment to Detect a Proton Electric Dipole Moment}, 
Rev. Sci. Instrum. {\bf 87}, 115116 (2016). %arXiv:1502.04317 [physics.acc-ph].

\bibitem{storage7}
T. Fukuyama, 
%\emph{General Spin Precession and Betatron Oscillation in Storage Ring}, 
Mod. Phys. Lett. A {\bf 31}, 1650135 (2016). 
%arXiv:1602.07923 [physics.acc-ph].

\bibitem{bnl}
Storage Ring EDM Collaboration [http://www.bnl.gov/edm/].

\bibitem{schiff}
L. I. Schiff, 
%\emph{Measurability of Nuclear Electric Dipole Moments}, 
Phys. Rev. {\bf 132}, 2194 (1963).

\bibitem{ckm}
M. Kobayashi and T. Maskawa, 
%\emph{CP Violation in the Renormalizable Theory of Weak Interaction}, 
Prog. Theor. Phys. {\bf 49}, 652 (1973).

\bibitem{smtheta}
I. B. Khriplovich, 
%\emph{Quark electric dipole moment and induced $\theta$ term in the Kobayashi-Maskawa model}, 
Phys. Lett. B {\bf 173}, 193 (1986); Yad. Fiz. {\bf 44}, 1019 (1986) [Sov. J. Nucl. Phys. {\bf 44}, 659 (1986)].

\bibitem{czarnecki}
A. Czarnecki and B. Krause, 
%\emph{Neutron electric dipole moment in the standard model: Valence quark contributions}, 
Phys. Rev. Lett. {\bf 78}, 4339 (1997).% [hep-ph/9704355].

\bibitem{smneutronedmmckellar}
B. McKellar, S. R. Choudhury, X.-G. He, and S. Pakvasa, 
%\emph{The neutron electric dipole moment in the Kobayashi-Maskawa model}, 
Phys. Lett. B {\bf 197}, 556 (1987).

\bibitem{mannel}
T. Mannel and N. Uraltsev, 
%\emph{Loop-less electric dipole moment of the nucleon in the standard model}, 
Phys. Rev. D {\bf 85}, 096002 (2012).% [arXiv:1202.6270].

\bibitem{seng}
C.-Y. Seng, 
%\emph{Reexamination of The Standard Model Nucleon Electric Dipole Moment}, 
Phys. Rev. C {\bf 91}, 025502 (2015).% [arXiv:1411.1476].

\bibitem{yamanakasmedm}
N. Yamanaka and E. Hiyama, 
%\emph{Standard model contribution to the electric dipole moment of the deuteron, $^3$H, and $^3$He nuclei}, 
J. High Energy Phys. 02 (2016) 067.% [arXiv:1512.03013].

\bibitem{sushkov}
O. P. Sushkov, V. V. Flambaum, and I. B. Khriplovich, 
%\emph{On the Possibility to Study P Odd and T Odd Nuclear Forces in Atomic and Molecular Experiments}, 
Zh. Eksp. Teor. Fiz. {\bf 87}, 1521 (1984) [Sov. Phys. JETP {\bf 60}, 873 (1984)].

\bibitem{sushkov2}
V. V. Flambaum, O. P. Sushkov, and I. B. Khriplovich, 
%\emph{Limit on the constant of T-nonconserving nucleon-nucleon interaction}, 
Phys. Lett. B {\bf 162}, 213 (1985).

\bibitem{sushkov3}
V. V. Flambaum, O. P. Sushkov, and I. B. Khriplovich, 
%\emph{On the P- and T-nonconserving nuclear moments}, 
Nucl. Phys. A {\bf 449}, 750 (1986).

\bibitem{avishai1}
Y. Avishai, 
%\emph{Electric dipole moment of the deuteron}, 
Phys. Rev. D {\bf 32}, 314 (1985).

\bibitem{avishai2}
Y. Avishai and M. Fabre de la Ripelle, 
%\emph{Electric Dipole Moment of $^3$He}, 
Phys. Rev. Lett. {\bf 56}, 2121 (1986).

\bibitem{avishai3}
Y. Avishai and M. Fabre de la Ripelle, 
%\emph{Electric dipole moment of $^3$He}, 
Nucl. Phys. A {\bf 468}, 578 (1987).

\bibitem{korkin}
I. B. Khriplovich and R. A. Korkin, 
%\emph{P and T odd electromagnetic moments of deuteron in chiral limit}, 
Nucl. Phys. A {\bf 665}, 365 (2000).% [nucl-th/9904081].

\bibitem{pospelovdeuteron}
O. Lebedev, K. A. Olive, M. Pospelov, and A. Ritz, 
%\emph{Probing CP violation with the deuteron electric dipole moment}, 
Phys. Rev. D {\bf 70}, 016003 (2004).% [hep-ph/0402023].

\bibitem{liu}
C.-P. Liu and R. G. E. Timmermans, 
%\emph{P- and T-odd two-nucleon interaction and the deuteron electric dipole moment}, 
Phys. Rev. C {\bf 70}, 055501 (2004).% [nucl-th/0408060].

\bibitem{afnan}
I. R. Afnan and B. F. Gibson, 
%\emph{Model dependence of the 2H electric dipole moment}, 
Phys. Rev. C {\bf 82}, 064002 (2010).% [arXiv:1011.4968].

\bibitem{devries}
J. de Vries, E. Mereghetti, R.G.E. Timmermans, and U. van Kolck, 
%\emph{Parity- and Time-Reversal-Violating Form Factors of the Deuteron}, 
Phys. Rev. Lett. {\bf 107}, 091804 (2011).% [arXiv:1102.4068].

\bibitem{dedmtheta}
J. Bsaisou, C. Hanhart, S. Liebig, U.-G. Mei{\ss}ner, A. Nogga, and A. Wirzba, 
%\emph{The electric dipole moment of the deuteron from the QCD $\theta$-term}, 
Eur. Phys. J. A {\bf 49}, 31 (2013).% [arXiv:1209.6306].

\bibitem{stetcu}
I. Stetcu, C.-P. Liu, J. L. Friar, A. C. Hayes, and P. Navratil, 
%\emph{Nuclear electric dipole moment of $^3$He}, 
Phys. Lett. B {\bf 665}, 168 (2008).% [arXiv:0804.3815].

\bibitem{chiral3nucleon}
J. de Vries, R. Higa, C.-P. Liu, E. Mereghetti, I. Stetcu, R. G. E. Timmermans, and U. van Kolck, 
%\emph{Electric Dipole Moments of Light Nuclei From Chiral Effective Field Theory}, 
Phys. Rev. C {\bf 84}, 065501 (2011).% [arXiv:1109.3604].

\bibitem{song}
Y.-H. Song, R. Lazauskas, and V. Gudkov, 
%\emph{Nuclear electric dipole moment of three-body systems}, 
Phys. Rev. C {\bf 87}, 015501 (2013).% [arXiv:1211.3762].

\bibitem{dekens}
W. Dekens, J. de Vries, J. Bsaisou, W. Bernreuther, C. Hanhart, U.-G. Mei{\ss}ner, A. Nogga, and A. Wirzba, 
%\emph{Unraveling models of CP-violation through electric dipole moments of light nuclei}, 
J. High Energy Phys. 07 (2014) 069.% [arXiv:1404.6082].

\bibitem{bsaisou}
J. Bsaisou, J. de Vries, C. Hanhart, S. Liebig, U.-G. Mei{\ss}ner, D. Minossi, A. Nogga, and A. Wirzba, 
%\emph{Nuclear Electric Dipole Moments in Chiral Effective Field Theory}, 
J. High Energy Phys. 03 (2015) 104 [Erratum ibid. 1505 (2015) 083].% [arXiv:1411.5804].

\bibitem{bsaisou2}
J. Bsaisou, U.-G. Mei{\ss}ner, A. Nogga, and A. Wirzba, 
%\emph{P- and T-Violating Lagrangians in Chiral Effective Field Theory and Nuclear Electric Dipole Moments}, 
Annals Phys. {\bf 359}, 317 (2015).% [arXiv:1412.5471].

\bibitem{mereghetti}
E. Mereghetti and U. van Kolck, 
%\emph{Effective Field Theory and Time-Reversal Violation in Light Nuclei}, 
Ann. Rev. Nucl. Part. Sci. {\bf 65}, 215 (2015).% [arXiv:1505.06272].

\bibitem{devries2}
J. de Vries, E. Mereghetti, and A. Walker-Loud, 
%\emph{Baryon mass splittings and strong CP violation in SU(3) Chiral Perturbation Theory}, 
Phys. Rev. C {\bf 92}, 045201 (2015).% [arXiv:1506.06247].

\bibitem{devries3}
J. de Vries and Ulf-G. Mei{\ss}ner, Int. J. Mod. Phys. E {\bf 25}, 1641008 (2016).
%\emph{Violations of discrete space-time symmetries in chiral effective field theory}, 
%arXiv:1509.07331 [hep-ph].

\bibitem{yamanakanuclearedm}
N. Yamanaka and E. Hiyama, 
%\emph{Enhancement of the CP-odd effect in the nuclear electric dipole moment of $^6$Li}, 
Phys. Rev. C {\bf 91}, 054005 (2015).% [arXiv:1503.04446].

\bibitem{yoshinaganuclearedm}
N. Yoshinaga, K. Higashiyama, R. Arai, and E. Teruya, 
%\emph{Nuclear electric dipole moments for the lowest 1/2$^+$ states in Xe and Ba isotopes}, 
Phys. Rev. C {\bf 89}, 045501 (2014).

\bibitem{clusterreview1}
W. von Oertzen, M. Freer, and Y. Kanada-En'yo, 
%\emph{Nuclear clusters and nuclear molecules}, 
Phys. Rep. {\bf 432}, 43 (2006).

\bibitem{clusterreview2}
H. Horiuchi, K. Ikeda, and K. Kato, 
%\emph{Recent Developments in Nuclear Cluster Physics}, 
Prog. Theor. Phys. Suppl. {\bf 192}, 1 (2011).

\bibitem{clusterreview3}
Y. Funaki, H. Horiuchi, and A. Tohsaki, 
%\emph{Cluster models from RGM to alpha condensation and beyond}, 
Prog. Part. Nucl. Phys. {\bf 82}, 78 (2015).

\bibitem{eft6dim}
J. de Vries, E. Mereghetti, R.G.E. Timmermans, and U. van Kolck, 
%\emph{The Effective Chiral Lagrangian From Dimension-Six Parity and Time-Reversal Violation},
Annals Phys. {\bf 338}, 50 (2013). % [arXiv:1212.0990].

\bibitem{yamada}
T. Yamada and Y. Funaki, 
%\emph{$\alpha$-cluster structures and monopole excitations in $^{13}$C}, 
Phys. Rev. C {\bf 92}, 034326 (2015). % [arXiv:1503.04261].

\bibitem{hiyama}
E. Hiyama, Y. Kino, and M. Kamimura, 
%\emph{Gaussian expansion method for few-body systems}, 
Prog. Part. Nucl. Phys. {\bf 51}, 223 (2003).

\bibitem{Hiyama2012ptep}
E. Hiyama, 
%\emph{Gaussian expansion method for few-body systems and its applications to atomic and nuclear physics}, 
Prog. Theor. Exp. Phys. {\bf 2012}, 01A204 (2012).

\bibitem{kamimuramuonic}
M. Kamimura, 
%\emph{Nonadiabatic coupled-rearrangement-channel approach to muonic molecules}, 
Phys. Rev. A {\bf 38}, 621 (1988).

\bibitem{3nucleon}
H. Kameyama, M. Kamimura, and Y. Fukushima, 
%\emph{Coupled-rearrangement-channel Gaussian-basis variational method for trinucleon bound states}, 
Phys. Rev. C {\bf 40}, 974 (1989).

\bibitem{hiyamahypernuclei1}
E. Hiyama, M. Kamimura, T. Motoba, T. Yamada, and Y. Yamamoto, 
%\emph{Three body model study of A = 6-7 hypernuclei: Halo and skin structures}, 
Phys. Rev. C {\bf 53}, 2075 (1996).
 
\bibitem{hiyamahypernuclei2}
E. Hiyama, M. Kamimura, T. Motoba, T. Yamada, and Y. Yamamoto, 
%\emph{$\Lambda N$ Spin-Orbit Splittings in $_\Lambda^9$Be and $_\Lambda^{13}$C Studied with One-Boson-Exchange $\Lambda N$ Interactions}, 
Phys. Rev. Lett. {\bf 85}, 270 (2000).

\bibitem{benchmark}
H. Kamada {\it et al.}, 
%\emph{Benchmark test calculation of a four nucleon bound state}, 
Phys. Rev. C {\bf 64}, 044001 (2001).% [nucl-th/0104057].

\bibitem{hiyamahyperon-nucleon}
E. Hiyama, K. Suzuki, H. Toki, and M. Kamimura, 
%\emph{Role of the Quark-Quark Correlation in Baryon Structure and Non-Leptonic Weak Transitions of Hyperons}, 
Prog. Theor. Phys. {\bf 112}, 99 (2004).% [nucl-th/0402007]. 

\bibitem{hiyama1stexcitedstate}
E. Hiyama, B. F. Gibson, and M. Kamimura, 
%\emph{Four-body calculation of the first excited state of $^4$He using a realistic $NN$ interaction: $^4$He $(e,e′)$ $^4$He$(0^+_2)$ and the monopole sum rule}, 
Phys. Rev. C {\bf 70}, 031001 (2004).

\bibitem{yamadaschuck}
T. Yamada and P. Schuck, Eur. Phys. J. A {\bf 26}, 185 (2005).

\bibitem{hiyamapentaquark}
E. Hiyama, M. Kamimura, A. Hosaka, H. Toki, and M. Yahiro, 
%\emph{Five-body calculation of resonance and scattering states of pentaquark system}, 
Phys. Lett. B {\bf 633}, 237 (2006).% [hep-ph/0507105].

\bibitem{hamaguchi}
K. Hamaguchi, T. Hatsuda, M. Kamimura, Y. Kino, and T. T. Yanagida, 
%\emph{Stau-catalyzed $^6$Li production in big-bang nucleosynthesis}, 
Phys. Lett. B {\bf 650}, 268 (2007).% [hep-ph/0702274].

\bibitem{funaki}
Y. Funaki, T. Yamada, H. Horiuchi, G. R\"{o}pke, P. Schuck and A. Tohsaki, 
%\emph{$\alpha$-Particle Condensation in $^{16}$O Studied with a Full Four-Body Orthogonality Condition Model Calculation}, 
Phys. Rev. Lett. {\bf 101}, 082502 (2008).% [arXiv:0802.3246].

\bibitem{hiyamahypernuclei3}
E. Hiyama, M. Kamimura, Y. Yamamoto, and T. Motoba, 
%\emph{Five-Body Cluster Structure of the Double-$\Lambda$ Hypernucleus $_{\Lambda \Lambda}^{11}$Be}, 
Phys. Rev. Lett. {\bf 104}, 212502 (2010). % [arXiv:1006.2626].

\bibitem{hiyamaatom1}
E. Hiyama and M. Kamimura, 
%\emph{Variational calculation of $^4$He tetramer ground and excited states using a realistic pair potential}, 
Phys. Rev. A {\bf 85}, 022502 (2012).% [arXiv:1111.4370].

\bibitem{ohtsubo}
S.-I. Ohtsubo, Y. Fukushima, M. Kamimura, and E. Hiyama, Prog. Theor. Exp. Phys. {\bf 2013}, 073D02 (2013).

\bibitem{yokota}
A. Yokota, E. Hiyama, and M. Oka, 
%\emph{Possible existence of charmonium–nucleus bound states}, 
Prog. Theor. Exp. Phys. {\bf 2013}, 113D01 (2013).% [arXiv:1308.6102].

\bibitem{hiyamaatom2}
E. Hiyama and M. Kamimura, 
%\emph{Universality in Efimov-associated tetramers in $^4$He}, 
Phys. Rev. A {\bf 90}, 052514 (2014).% [arXiv:1409.2501].

\bibitem{kusakabe}
M. Kusakabe, K. S. Kim, M.-K. Cheoun, T. Kajino, Y. Kino, and G. J. Mathews, 
%\emph{Revised Big Bang Nucleosynthesis with long-lived negatively charged massive particles: updated recombination rates, primordial $^9$Be nucleosynthesis, and impact of new $^6$Li limits}, 
Astrophys. J. Suppl. {\bf 214}, 5 (2014).% [arXiv:1403.4156].

\bibitem{maeda}
S. Maeda, M. Oka, A. Yokota, E. Hiyama, and Y.-R. Liu, 
%\emph{A model of charmed baryon-nucleon potential and 2- and 3-body bound states with charmed baryon}, 
%arXiv:1509.02445 [nucl-th].
Prog. Theor. Exp. Phys. (2016) 023D02.

\bibitem{schmid}
E. W. Schmid and K. Wildermuth, 
%\emph{Phase Shift Calculations on $\alpha - \alpha$ Scattering}, 
Nucl. Phys. {\bf 26}, 463 (1961).

\bibitem{kanada}
H. Kanada, T. Kaneko, S. Nagata, and M. Morikazu, 
%\emph{Microscopic Study of Nucleon-$^4$He Scattering and Effective Nuclear Potentials}, 
Prog. Theor. Phys. {\bf 61}, 1327 (1979).

\bibitem{freer}
M. Freer {\it et al.}, 
%\emph{Evidence for a new $^{12}$C state at 13.3 MeV}, 
Phys. Rev. C {\bf 83}, 034314 (2011).

\bibitem{itoh}
M. Itoh {\it et al.}, 
%\emph{Candidate for the 2$^+$ excited Hoyle state at E$_x \sim $10 MeV in $^{12}$C}, 
Phys. Rev. C {\bf 84}, 054308 (2011).

\bibitem{ocm1}
S. Saito, 
%\emph{Effect of Pauli Principle in Scatterings of Two Clusters}, 
Prog. Theor. Phys. {\bf 40}, 893 (1968).

\bibitem{ocm2}
S. Saito, 
%\emph{Interaction between Clusters and Pauli Principle}, 
Prog. Theor. Phys. {\bf 41}, 705 (1969).

\bibitem{horiuchi1}
H. Horiuchi, 
%\emph{Multi-Cluster Allowed States and Spectroscopic Amplitude of Cluster Transfer}, 
Prog. Theor. Phys. {\bf 58}, 204 (1977).

\bibitem{ocm3}
S. Saito, 
%\emph{Chapter II. Theory of Resonating Group Method and Generator Coordinate Method, and Orthogonality Condition Model}, 
Prog. Theor. Phys. Suppl. No. {\bf 62}, 11 (1977).

\bibitem{horiuchi2}
H. Horiuchi, \emph{Chapter III. Kernels of GCM, RGM and OCM and their calculational methods}, Prog. Theor. Phys. Suppl. {\bf 62}, 90 (1977).

\bibitem{Kukulin}
V. I. Kukulin, V. N. Pomerantsev, Kh. D. Razikov, V. T. Voronchev, and G. G. Ryzhinkh, 
%\emph{Detailed study of the cluster structure of light nuclei in a three-body model (IV). Large space calculation for A = 6 nuclei with realistic nuclear forces}, 
Nucl. Phys. A {\bf 586}, 151 (1995).

\bibitem{crewther}
R. J. Crewther, P. Di Vecchia, G. Veneziano, and E. Witten, 
%\emph{Chiral estimate of the electric dipole moment of the neutron in quantum chromodynamics}, 
Phys. Lett. B {\bf 88}, 123 (1979) [Erratum ibid. B {\bf 91}, 487 (1980)].

\bibitem{Barton}
G. Barton and E. G.White, Phys. Rev. {\bf 184}, 1660 (1969).

\bibitem{pvcpvhamiltonian1}
W. C. Haxton and E. M. Henley, 
%\emph{Enhanced T-nonconserving nuclear moments}, 
Phys. Rev. Lett. {\bf 51}, 1937 (1983).

\bibitem{pvcpvhamiltonian2}
V. P. Gudkov, X.-G. He, and B. H. J. McKellar, 
%\emph{CP-odd nucleon potential}, 
Phys. Rev. C {\bf 47}, 2365 (1993).% [hep-ph/9212207].

\bibitem{pvcpvhamiltonian3}
I. S. Towner and A. C. Hayes, 
%\emph{P, T-violating nuclear matrixx elements in the one-meson exchange approximation}, 
Phys. Rev. C {\bf 49}, 2391 (1994).% [nucl-th/9402026].

\bibitem{sandars1}
P. G. H. Sandars, 
%\emph{The electric dipole moment of an atom}, 
Phys. Lett. {\bf 14}, 194 (1965).

\bibitem{sandars2}
P. G. H. Sandars, 
%\emph{Enhancement factor for the electric dipole moment of the valence electron in an alkali atom}, 
Phys. Lett. {\bf 22}, 290 (1966).

\bibitem{flambaumenhancement}
V. V. Flambaum, 
%\emph{On enhancement of the electron electric dipole moments in heavy atoms}, 
Yad. Fiz. {\bf 24}, 383 (1976) [Sov. J. Nucl. Phys. {\bf 24}, 199 (1976)].

\bibitem{inoue}
S. Inoue, V. Gudkov, M. R. Schindler, and Y.-H. Song, Phys. Rev. C {\bf 93}, 055501 (2016).
%\emph{Screening of Nucleon Electric Dipole Moments in Nuclei}, arXiv:1512.06131 [nucl-th].

\bibitem{kurokawa}
C. Kurokawa and K. Kato, Phys. Rev. C {\bf 71}, 021301 (2005). 

\bibitem{pastore1}
S. Pastore, R. Schiavilla, and J. L. Goity, 
%\emph{Electromagnetic two-body currents of one- and two-pion range}, 
Phys. Rev. C {\bf 78}, 064002 (2008).% [arXiv:0810.1941].

\bibitem{pastore2}
S. Pastore, L. Girlanda, R. Schiavilla, M. Viviani, and R. B. Wiringa, 
%\emph{Electromagnetic currents and magnetic moments in chiral effective field theory ($\chi$EFT)}, 
Phys. Rev. C {\bf 80}, 034004 (2009).% [arXiv:0906.1800].

\bibitem{pastore3}
S. Pastore, L. Girlanda, R. Schiavilla, and M. Viviani, 
%\emph{Two-nucleon electromagnetic charge operator in chiral effective field theory ($\chi$EFT) up to one loop}, 
Phys. Rev. C {\bf 84}, 024001 (2011).% [arXiv:1106.4539].

\bibitem{pastore4}
S. Pastore, Steven C. Pieper, R. Schiavilla, and R. B. Wiringa, 
%\emph{Quantum Monte Carlo calculations of electromagnetic moments and transitions in A $\leq$ 9 nuclei with meson-exchange currents derived from chiral effective field theory}, 
Phys. Rev. C {\bf 87}, 035503 (2013).% [arXiv:1212.3375].

\bibitem{ibrahim}
T. Ibrahim and P. Nath, 
%\emph{The neutron and the lepton EDMs in MSSM, large CP-violating phases and the cancellation mechanism}, 
Phys. Rev. D {\bf 58}, 111301 (1998) [Erratum ibid. D {\bf 60}, 099902 (1999)]. % [hep-ph/9807501].

\bibitem{ramseyli}
Y. Li, S. Profumo, and M. Ramsey-Musolf, 
%\emph{A comprehensive analysis of electric dipole moment constraints on CP-violating phases in the MSSM}, 
J. High Energy Phys. 08 (2010) 062.% [arXiv:1006.1440].

\bibitem{ellisgeometric}
J. Ellis, J. S. Lee, and A. Pilaftsis, 
%\emph{A geometric approach to CP-violation: applications to the MCPMFV SUSY model}, 
J. High Energy Phys. 10 (2010) 049.% [arXiv:1006.3087].

\bibitem{ellisgeometric2}
J. Ellis, J. S. Lee, and A. Pilaftsis, 
%\emph{Maximal electric dipole moments of nuclei with enhanced Schiff moments}, 
J. High Energy Phys. 02 (2011) 045.% [arXiv:1101.3529].

\bibitem{rpvlinearprogramming}
N. Yamanaka, T. Sato, and T. Kubota, 
%\emph{Linear programming analysis of the R-parity violation within EDM-constraints}, 
J. High Energy Phys. 12 (2014) 110.% [arXiv:1406.3713].

\bibitem{ellismssmedm}
J. R. Ellis, S. Ferrera, and D. V. Nanopoulos, Phys. Lett. B {\bf 114}, 231 (1982).

\bibitem{chang}
D. Chang, W.-Y. Keung, and A. Pilaftsis, Phys. Rev. Lett. {\bf 82}, 900 (1999).

\bibitem{demir}
D. Demir, O. Lebedev, K. A. Olive, M. Pospelov, and A. Ritz, 
%\emph{Electric dipole moments in the MSSM at large tan beta}, 
Nucl. Phys. B {\bf 680}, 339 (2004).% [hep-ph/0311314].

\bibitem{mssmreloaded}
J. R. Ellis, J. S. Lee and A. Pilaftsis, 
%\emph{Electric dipole moments in the MSSM reloaded}, 
J. High Energy Phys. 10 (2008) 049.% [arXiv:0808.1819].

\bibitem{mssmrainbow}
N. Yamanaka, 
%\emph{Two-loop level rainbowlike supersymmetric contribution to the fermion electric dipole moment}, 
Phys. Rev. D {\bf 87}, 011701 (2013).% [arXiv:1211.1808].

\bibitem{barr-zee}
S. M. Barr and A. Zee, 
%\emph{Electric Dipole Moment of the Electron and of the Neutron}, 
Phys. Rev. Lett. {\bf 65}, 21 (1990).

\bibitem{jung2hdmedm}
M. Jung and A. Pich, J. High Energy Phys. 04 (2014) 076.

\bibitem{abe}
T. Abe, J. Hisano, T. Kitahara, and K. Tobioka, 
%\emph{Gauge invariant Barr-Zee type contributions to fermionic EDMs in the two-Higgs doublet models}, 
J. High Energy Phys. 01 (2014) 106.% [arXiv:1311.4704].

\bibitem{rpv1}
N. Yamanaka, T. Sato, and T. Kubota, 
%\emph{Reappraisal of two-loop contributions to the fermion electric dipole moments in R-parity violating supersymmetric model}, 
Phys. Rev. D {\bf 85}, 117701 (2012).% [arXiv:1202.0106].

\bibitem{rpv2}
N. Yamanaka, 
%\emph{Sfermion loop contribution to the two-loop level fermion electric dipole moment in R-parity violating supersymmetric models}, 
Phys. Rev. D {\bf 86}, 075029 (2012).% [arXiv:1208.4521].

\bibitem{rpv3}
N. Yamanaka, T. Sato, and T. Kubota, 
%\emph{R-parity violating supersymmetric Barr-Zee type contributions to the fermion electric dipole moment with weak gauge boson exchange}, 
Phys. Rev. D {\bf 87}, 115011 (2013).% [arXiv:1212.6833].

\bibitem{rpv4}
N. Yamanaka, 
%\emph{R-parity violating two-loop level rainbowlike contribution to the fermion electric dipole moment}, 
arXiv:1212.5800 [hep-ph].

\bibitem{bacchetta}
A. Bacchetta, A. Courtoy, and M. Radici, 
%\emph{First extraction of valence transversities in a collinear framework}, 
J. High Energy Phys. 03 (2013) 119. % [arXiv:1212.3568].

\bibitem{anselmino}
M. Anselmino, M. Boglione, U. D'Alesio, S. Melis, F. Murgia, and A. Prokudin, 
%\emph{Simultaneous extraction of transversity and Collins functions from new semi-inclusive deep inelastic scattering and $e^+e^-$ data}, 
Phys. Rev. D {\bf 87}, 094019 (2013).% [arXiv:1303.3822].

\bibitem{courtoy}
A. Courtoy, S. Bae{\ss}ler, M. Gonz\'{a}lez-Alonso, and S. Liuti, 
%\emph{Beyond-Standard-Model Tensor Interaction and Hadron Phenomenology}, 
Phys. Rev. Lett. {\bf 115}, 162001 (2015).% [arXiv:1503.06814].

\bibitem{kang}
Z.-B. Kang, A. Prokudin, P. Sun, and F. Yuan, 
%\emph{Extraction of quark transversity distribution and Collins fragmentation functions with QCD evolution}, 
Phys. Rev. D {\bf 93}, 014009 (2016).% [arXiv:1505.05589].

\bibitem{saoki}
S. Aoki, M. Doui, T. Hatsuda, and Y. Kuramashi, 
%\emph{Tensor charge of the nucleon in lattice QCD}, 
Phys. Rev. D {\bf 56}, 433 (1997).% [hep-lat/9608115].

\bibitem{green}
J. R. Green {\it et al.}, 
%\emph{Nucleon scalar and tensor charges from lattice QCD with light Wilson quarks}, 
Phys. Rev. D {\bf 86}, 114509 (2012).% [arXiv:1206.4527].

\bibitem{yaoki}
Y. Aoki {\it et al.} (UKQCD Collaboration), 
%\emph{Nucleon isovector structure functions in (2+1)-flavor QCD with domain wall fermions}, 
Phys. Rev. D {\bf 82}, 014501 (2010).% [arXiv:1003.3387].

\bibitem{bhattacharya1}
T. Bhattacharya, S. D. Cohen, R. Gupta, A. Joseph, H.-W. Lin, and B. Yoon, 
%\emph{Nucleon charges and electromagnetic form factors from 2+1+1-flavor lattice QCD}, 
Phys. Rev. D {\bf 89}, 094502 (2014).% [arXiv:1306.5435].

\bibitem{etm2}
A. Abdel-Rehim {\it et al.}, 
%\emph{Disconnected quark loop contributions to nucleon observables in lattice QCD}, 
Phys. Rev. D {\bf 89}, 034501 (2014).% [arXiv:1310.6339].

\bibitem{bhattacharya2}
T. Bhattacharya {\it et al.}, 
%\emph{Neutron Electric Dipole Moment and Tensor Charges from Lattice QCD}, 
Phys. Rev. Lett. {\bf 115}, 212002 (2015).% [arXiv:1506.04196].

\bibitem{bhattacharya3}
T. Bhattacharya {\it et al.}, 
%\emph{Isovector and Iso-scalar Tensor Charges of the Nucleon from Lattice QCD}, 
Phys. Rev. D {\bf 92}, 094511 (2015).% [arXiv:1506.06411].

\bibitem{etm3}
A. Abdel-Rehim {\it et al.}, 
%\emph{Nucleon and pion structure with lattice QCD simulations at physical value of the pion mass}, 
Phys. Rev. D {\bf 92}, 114513 (2015).% [arXiv:1507.04936].

\bibitem{jlqcd4}
N. Yamanaka, H. Ohki, S. Hashimoto, and T. Kaneko (JLQCD Collaboration), 
%\emph{Nucleon axial and tensor charges with dynamical overlap quarks}, 
PoS LATTICE2015, 121 (2016).
%arXiv:1511.04589 [hep-lat].

\bibitem{yamanakasde1}
N. Yamanaka, T. M. Doi, S. Imai, and H. Suganuma, 
%\emph{Quark tensor charge and electric dipole moment within the Schwinger-Dyson formalism}, 
Phys. Rev. D {\bf 88}, 074036 (2013).% [arXiv:1307.4208].

\bibitem{yamanakasde2}
N. Yamanaka, S. Imai, T. M. Doi, and H. Suganuma, 
%\emph{Quark scalar, axial, and pseudoscalar charges in the Schwinger-Dyson formalism}, 
Phys. Rev. D {\bf 89}, 074017 (2014).% [arXiv:1401.2852].

\bibitem{pitschmann}
M. Pitschmann, C.-Y. Seng, C. D. Roberts, and S. M. Schmidt, 
%\emph{Nucleon tensor charges and electric dipole moments}, 
Phys. Rev. D {\bf 91}, 074004 (2015). %[arXiv:1411.2052].

\bibitem{tensorrenormalization}
V. Barone, Phys. Lett. B {\bf 409}, 499 (1997); 
X. Artru and M. Mekhfi, Z. Phys. C {\bf 45}, 669 (1990).

\bibitem{degrassi}
G. Degrassi, S. Marchetti, E. Franco, and L. Silvestrini, 
%\emph{QCD corrections to the electric dipole moment of the neutron in the MSSM}, 
J. High Energy Phys. 11 (2005) 044. % [hep-ph/0510137].

\bibitem{yang}
J. Hisano, K. Tsumura, and M. J. S. Yang, 
%\emph{QCD corrections to neutron electric dipole moment from dimension-six four-quark operators}, 
Phys. Lett. B {\bf 713}, 473 (2012).% [arXiv:1205.2212].


\end{thebibliography}
\end{document}